\documentclass[10pt,journal,twoside]{IEEEtran}
\usepackage{amsmath}
\usepackage{amsthm}
\usepackage{amsfonts}
\usepackage{graphicx}
\usepackage{latexsym}
\usepackage{amssymb}
\usepackage{color}
\usepackage{url}

\IfFileExists{slashbox.sty}{\usepackage{slashbox}}{} % Not used below.
\usepackage{xfrac}

\usepackage{array}
\DeclareMathAlphabet{\mathbfsl}{OT1}{ppl}{b}{it} %{OT1}{cmr}{bx}{it}

\makeatletter
\DeclareRobustCommand{\nsbinom}{\genfrac[]\z@{}}
\makeatother

\newcommand{\field}[1]{\mathbb{#1}}

\newcommand{\F}{\field{F}}

\newcommand{\cA}{{\cal A}}

\newcommand{\linadd}{\kern1pt\mbox{\small$\boxplus$}\kern1pt}

\newtheorem{theorem}{Theorem}

\newtheorem{conjecture}{Conjecture}

\theoremstyle{definition}

\newcommand{\old}[1]{\hspace{1.5pt}#1\hspace{1.5pt}}
\newcounter{diagramrowcount}
\newcommand{\diagramrow}[2]{%
  \stepcounter{diagramrowcount}%
  \par\nointerlineskip
  \vbox to68pt{\vfil
    \hbox to\linewidth{\hfil$\displaystyle
      \ifnum\value{diagramrowcount}=1
        \left[\begin{array}{@{}#1@{}|}#2\end{array}\right.
      \else\ifnum\value{diagramrowcount}=9
        \left.\begin{array}{@{}#1@{}}#2\end{array}\right]
      \else
        \begin{array}{@{}#1@{}|}#2\end{array}
      \fi\fi
    $\hfil}%
  \vfil}}

\newcommand{\parityheader}[1]{%
  \setcounter{diagramrowcount}{0}%
  \hbox to\linewidth{\hfil{\fontsize{12}{14}\selectfont\textbf{#1}}\hfil}%
  \vspace{2pt}}

\newcommand{\seedtitle}[1]{%
  \par\nointerlineskip
  \hbox to\linewidth{\hfil\textbf{#1}\hfil}%
  \nobreak\vskip5pt}
\newcommand{\seedstrip}[5]{%
  \par\nointerlineskip
  \vbox{%
    \hbox to\linewidth{\hfil#1\hfil}%
    \kern2pt
    \hbox to\linewidth{\hfil$\displaystyle
      \left#3\begin{array}{@{}*{#2}{c}@{}}#5\end{array}\right#4
    $\hfil}}%
  \vskip4pt}

\begin{document}

\bibliographystyle{plain}

\title{Ternary Perfect $2$-Burst-Correcting Codes}

\author{\IEEEauthorblockN{Tuvi Etzion}
\IEEEauthorblockA{\textit{Dept. of Computer Science, Technion-Israel Institute of Technology} \\
%\textit{Technion-Israel Institute of Technology}\\
Haifa 3200003, Israel (etzion@cs.technion.ac.il)}
}

\maketitle
\thispagestyle{plain}

\begin{abstract}
A construction for a ternary perfect linear $2$-burst-correcting code is presented
for every redundancy $r \geq 5$.
The construction is recursive with seeds given for $5 \leq r \leq 7$. The construction of
redundancy $r \geq 8$ is derived from a code of redundancy $r-3$ and it
distinguishes between codes of even and odd length.
\end{abstract}


\section{Introduction}

Detecting and correcting errors in memory systems is of great importance.
Background on errors in memories such as byte-oriented codes, memory
applications, and error control is extensively studied in~\cite{RaFu89}.
Further such work, for example on byte-correcting codes, can be found in~\cite{Che83,Che86,Etz98}.
A burst of errors is very common in such memories.
In this paper, we consider ternary perfect $2$-burst-correcting codes.

We first recall the classical bounds for linear codes. Let
$(n,k)$ code denote a linear code of length $n$ and dimension $k$. The
redundancy $r$ of such a code is $r=n-k$. A code is called a
\emph{$b$-burst-correcting code} if it can correct any single burst of length
$b$ or less. A code is called \emph{cyclic $b$-burst-correcting code} if it can
correct any single cyclic burst of length $b$ or less. Reiger~\cite{Rei60} has proved that an $(n,n-r)$
$b$-burst-correcting code (or cyclic $b$-burst-correcting code), where $n \geq 2b$, must satisfy $r \geq 2b$.

For an $(n,n-r)$ cyclic $b$-burst-correcting code over $\F_q$, the finite field with $q$ elements,
there are $n(q-1)q^{b-1}$ different cyclic bursts of length up to $b$, each one corresponding to a
different nonzero syndrome. Since the number of nonzero vectors of length
$r$ is $q^r-1$, it follows that $q^r-1\geq n(q-1)q^{b-1}$. Abramson~\cite{Abr59}
noted that this inequality for binary codes, and along with the fact that $n$ is an integer, it implies that
\begin{equation}\label{eq:cyclic}
n\leq (q^{r-b+1}-1)/(q-1).
\end{equation}

A cyclic $b$-burst-correcting code which satisfies \eqref{eq:cyclic} with
equality is said to be \emph{optimum}. Abramson~\cite{Abr59} has constructed
optimum binary cyclic $2$-burst-correcting codes for any $r\geq 3$. Elspas and
Short~\cite{ElSh62} constructed such codes for $b=3$ and $b=4$, Abdel-Ghaffar,
McEliece, Odlyzko, and van Tilborg~\cite{abdel} have shown that infinitely
many optimum codes exist for each $b\geq 1$. It is clear that if a code
satisfies \eqref{eq:cyclic} with equality then $q^r-1>n(q-1)q^{b-1}$ (when
$b>1$), i.e., not all nonzero vectors of length $r$ are syndromes obtained
as sums of at most $b$ adjacent columns. Therefore, these optimum codes
are not perfect. For an (acyclic) $(n,n-r)$ $b$-burst-correcting code, there are
$q^{b-1} ((q-1)(n-b+1)+1)-1$ different bursts of length up to $b$, each one
corresponding to a different nonzero syndrome. Again, since the number of
nonzero vectors of length $r$ is $q^r-1$, it follows that
$q^r-1\geq q^{b-1} ((q-1)(n-b+1)+1)-1$, i.e.,
\begin{equation}
\label{eq:perfect}
n\leq (q^{r-b+1}-1)/(q-1)+b-1.
\end{equation}

A $b$-burst-correcting code which satisfies \eqref{eq:perfect} with equality
is said to be \emph{perfect} (or ``complete'' as called by Reiger~\cite{Rei60}).

Abdel-Ghaffar~\cite{abdel-qary} extended the study of optimum cyclic
burst-correcting codes to $\F_q$. He gave a necessary polynomial
condition for attaining the length
$n=(q^{r-b+1}-1)/(q-1)$, and proved that, whenever this condition holds,
infinitely many such optimum codes exist.

In~\cite{Etz01} perfect linear binary $2$-burst-correcting codes
of length $2^{r-1}$ and redundancy $r$, were constructed for every $r\geq 5$.
The parity-check formulation requires each nonzero syndrome to occur exactly
once among the columns and the sums of adjacent columns.
The recursive construction in~\cite{Etz01}
replaces an $r \times 2^{r-1}$ parity-check matrix $H=[h_1,\ldots,h_n]$, where $n=2^{r-1}$, by the
$(r+2)\times 4n$ parity-check matrix $\widetilde H$ shown in
Fig.~\ref{fig:binary-recursion}.
\begin{figure*}[!t]
\begingroup
\setlength{\arraycolsep}{0.75pt}
\medmuskip=2mu
\renewcommand{\arraystretch}{1.15}
\[
\left[\begin{array}{*{5}{c}|*{5}{c}|*{5}{c}|*{5}{c}}
h_1 & h_2 & \cdots & h_{n-1} & h_n &
h_n & h_{n-1} & \cdots & h_2 & h_2+h_1 &
h_1+h_2 & h_2 & \cdots & h_{n-1} & h_{n-1}+h_n &
h_n+h_{n-1} & h_{n-1} & \cdots & h_2 & h_1 \\
1 & 1 & \cdots & 1 & 1 &
0 & 0 & \cdots & 0 & 0 &
1 & 0 & \cdots & 1 & 1 &
1 & 0 & \cdots & 1 & 0 \\
0 & 1 & \cdots & 0 & 1 &
0 & 0 & \cdots & 0 & 0 &
0 & 1 & \cdots & 1 & 0 &
1 & 1 & \cdots & 0 & 1
\end{array}\right].
\]
\endgroup
\caption{Binary recursion from \cite[p.~2554]{Etz01}. Vertical lines separate the four blocks.}
\label{fig:binary-recursion}
\end{figure*}

%FIRST TO CHANGE

\begin{figure*}[!t]
\centering
\scalebox{0.82}{\vbox{%
\begingroup
\fontsize{12}{12}\selectfont
\setlength{\arraycolsep}{0pt}
\renewcommand{\arraystretch}{1.0}
\seedtitle{Seed for \(r=5\): \(5\times 41\) matrix}

\seedstrip{Columns 1--41}{41}{[}{]}{%
  0 & 2 & 2 & 0 & 1 & 2 & 1 & 2 & 0 & 1 & 1 & 2 & 2 & 2 & 0 & 1 & 0 & 0 & 1 & 2 & 0 & 0 & 2 & 0 & 0 & 0 & 1 & 1 & 1 & 0 & 2 & 2 & 1 & 2 & 2 & 0 & 1 & 1 & 2 & 2 & 1\\
  1 & 2 & 1 & 1 & 1 & 2 & 2 & 2 & 0 & 0 & 0 & 0 & 2 & 1 & 0 & 2 & 2 & 1 & 0 & 2 & 1 & 2 & 2 & 2 & 1 & 0 & 1 & 0 & 1 & 0 & 1 & 0 & 1 & 2 & 2 & 1 & 1 & 0 & 2 & 0 & 0\\
  1 & 0 & 0 & 1 & 2 & 1 & 1 & 0 & 1 & 2 & 1 & 1 & 2 & 0 & 2 & 2 & 2 & 2 & 1 & 1 & 2 & 0 & 2 & 0 & 0 & 0 & 0 & 2 & 1 & 0 & 1 & 2 & 0 & 2 & 0 & 1 & 1 & 2 & 1 & 2 & 0\\
  1 & 1 & 2 & 1 & 1 & 1 & 0 & 0 & 1 & 1 & 1 & 1 & 0 & 1 & 1 & 1 & 2 & 1 & 0 & 1 & 2 & 2 & 1 & 0 & 0 & 2 & 2 & 0 & 0 & 1 & 1 & 2 & 1 & 2 & 1 & 0 & 2 & 2 & 2 & 1 & 1\\
  1 & 1 & 1 & 0 & 0 & 0 & 1 & 1 & 1 & 0 & 0 & 0 & 1 & 1 & 0 & 0 & 1 & 0 & 1 & 1 & 1 & 1 & 0 & 1 & 1 & 1 & 1 & 1 & 0 & 1 & 0 & 1 & 0 & 1 & 0 & 1 & 1 & 1 & 1 & 0 & 1%
}

\par

\seedtitle{Seed for \(r=6\): \(6\times 122\) matrix}

\seedstrip{Columns 1--61}{61}{[}{.}{%
  2 & 2 & 2 & 0 & 2 & 0 & 2 & 0 & 0 & 0 & 1 & 0 & 2 & 1 & 2 & 0 & 2 & 1 & 2 & 1 & 1 & 0 & 2 & 1 & 2 & 1 & 0 & 1 & 0 & 1 & 2 & 0 & 0 & 2 & 1 & 2 & 2 & 1 & 2 & 0 & 1 & 1 & 1 & 1 & 0 & 0 & 1 & 1 & 0 & 2 & 0 & 2 & 0 & 1 & 2 & 1 & 2 & 1 & 0 & 0 & 2\\
  1 & 1 & 1 & 1 & 0 & 2 & 0 & 2 & 0 & 2 & 0 & 2 & 1 & 1 & 2 & 2 & 2 & 0 & 2 & 1 & 1 & 1 & 1 & 2 & 0 & 2 & 0 & 2 & 1 & 0 & 2 & 1 & 2 & 0 & 2 & 2 & 1 & 0 & 2 & 1 & 1 & 0 & 2 & 1 & 0 & 0 & 0 & 1 & 1 & 2 & 1 & 1 & 1 & 2 & 0 & 0 & 0 & 1 & 0 & 0 & 2\\
  2 & 0 & 2 & 1 & 1 & 1 & 2 & 2 & 2 & 0 & 2 & 1 & 0 & 0 & 0 & 0 & 2 & 2 & 0 & 2 & 1 & 0 & 0 & 2 & 2 & 2 & 0 & 0 & 1 & 0 & 2 & 2 & 2 & 0 & 1 & 2 & 0 & 2 & 2 & 2 & 1 & 1 & 1 & 2 & 0 & 0 & 0 & 1 & 0 & 1 & 2 & 2 & 2 & 2 & 1 & 1 & 2 & 1 & 2 & 1 & 1\\
  1 & 1 & 2 & 1 & 1 & 1 & 1 & 1 & 0 & 1 & 0 & 1 & 2 & 2 & 2 & 2 & 2 & 1 & 1 & 1 & 2 & 1 & 2 & 2 & 1 & 2 & 0 & 2 & 1 & 1 & 2 & 1 & 1 & 1 & 2 & 0 & 2 & 1 & 1 & 1 & 2 & 1 & 2 & 2 & 2 & 1 & 1 & 2 & 1 & 1 & 1 & 2 & 1 & 2 & 1 & 1 & 1 & 2 & 1 & 0 & 2\\
  1 & 1 & 2 & 1 & 1 & 2 & 0 & 0 & 2 & 2 & 1 & 0 & 1 & 0 & 1 & 1 & 0 & 0 & 1 & 0 & 2 & 1 & 0 & 0 & 0 & 1 & 0 & 0 & 2 & 0 & 1 & 0 & 1 & 1 & 0 & 1 & 2 & 2 & 0 & 2 & 0 & 2 & 2 & 1 & 1 & 1 & 1 & 1 & 2 & 1 & 0 & 1 & 1 & 2 & 0 & 0 & 1 & 1 & 0 & 0 & 1\\
  0 & 0 & 1 & 1 & 1 & 1 & 1 & 1 & 1 & 1 & 0 & 1 & 1 & 1 & 1 & 0 & 1 & 1 & 0 & 0 & 1 & 1 & 1 & 1 & 0 & 1 & 1 & 1 & 1 & 1 & 1 & 0 & 1 & 1 & 1 & 0 & 1 & 1 & 0 & 1 & 1 & 1 & 1 & 1 & 0 & 0 & 1 & 0 & 1 & 0 & 1 & 1 & 1 & 1 & 0 & 0 & 1 & 1 & 0 & 1 & 1%
}

\seedstrip{Columns 62--122}{61}{.}{]}{%
  2 & 2 & 2 & 1 & 0 & 1 & 1 & 0 & 2 & 0 & 2 & 1 & 2 & 1 & 2 & 1 & 1 & 2 & 0 & 0 & 1 & 1 & 1 & 1 & 2 & 1 & 2 & 0 & 1 & 0 & 2 & 0 & 0 & 2 & 0 & 0 & 0 & 1 & 2 & 1 & 2 & 2 & 1 & 0 & 0 & 0 & 1 & 1 & 2 & 0 & 2 & 0 & 2 & 2 & 1 & 2 & 2 & 2 & 0 & 0 & 0\\
  0 & 0 & 1 & 1 & 1 & 1 & 2 & 2 & 2 & 2 & 2 & 0 & 2 & 2 & 2 & 2 & 2 & 2 & 0 & 2 & 1 & 1 & 0 & 1 & 0 & 0 & 0 & 2 & 0 & 0 & 2 & 0 & 0 & 1 & 0 & 1 & 0 & 0 & 0 & 0 & 2 & 2 & 1 & 1 & 0 & 2 & 2 & 0 & 2 & 1 & 1 & 1 & 2 & 1 & 1 & 1 & 1 & 0 & 2 & 1 & 2\\
  2 & 0 & 1 & 2 & 2 & 0 & 1 & 2 & 1 & 2 & 2 & 1 & 1 & 1 & 0 & 0 & 2 & 0 & 1 & 1 & 2 & 0 & 2 & 0 & 1 & 2 & 0 & 0 & 1 & 2 & 1 & 1 & 0 & 1 & 2 & 0 & 0 & 0 & 2 & 1 & 0 & 2 & 0 & 0 & 1 & 0 & 0 & 2 & 1 & 1 & 0 & 1 & 0 & 2 & 1 & 1 & 1 & 1 & 0 & 1 & 1\\
  2 & 1 & 2 & 2 & 2 & 2 & 2 & 1 & 2 & 2 & 2 & 1 & 2 & 0 & 2 & 2 & 1 & 0 & 0 & 1 & 2 & 1 & 2 & 2 & 1 & 1 & 1 & 1 & 0 & 0 & 2 & 1 & 0 & 2 & 0 & 0 & 0 & 1 & 1 & 1 & 1 & 2 & 2 & 1 & 0 & 1 & 2 & 1 & 1 & 1 & 2 & 1 & 2 & 2 & 0 & 1 & 2 & 1 & 0 & 0 & 0\\
  1 & 0 & 1 & 2 & 1 & 1 & 1 & 2 & 1 & 1 & 2 & 1 & 2 & 1 & 0 & 1 & 0 & 1 & 2 & 1 & 0 & 0 & 1 & 2 & 0 & 0 & 2 & 1 & 1 & 1 & 0 & 1 & 2 & 0 & 0 & 1 & 1 & 2 & 2 & 0 & 0 & 1 & 1 & 0 & 1 & 0 & 2 & 1 & 0 & 0 & 1 & 0 & 2 & 0 & 0 & 0 & 2 & 2 & 1 & 1 & 1\\
  0 & 1 & 1 & 1 & 0 & 1 & 1 & 1 & 0 & 0 & 1 & 1 & 1 & 0 & 1 & 1 & 0 & 0 & 1 & 1 & 1 & 0 & 0 & 1 & 1 & 0 & 1 & 1 & 0 & 1 & 1 & 0 & 1 & 1 & 1 & 0 & 1 & 1 & 1 & 1 & 0 & 0 & 0 & 1 & 1 & 1 & 1 & 1 & 0 & 1 & 0 & 0 & 1 & 1 & 0 & 0 & 1 & 1 & 0 & 0 & 0%
}
\endgroup
}}\par
\caption{Seed parity-check matrices for $r=5$ and $r=6$.}
\label{fig:seeds-five-six}
\end{figure*}

There are other related works in the literature.
Wei and Schwartz~\cite{wei-schwartz} studied perfect codes for bursts of
limited-magnitude errors through lattice tilings and splittings of finite
Abelian groups. They constructed families for cyclic and non-cyclic bursts
of length two and gave further constructions based on finite fields.
Sun, Lu, Zhang, and Ge~\cite{sun-burst} considered a $(t,s)$-burst, which
replaces $t$ consecutive symbols by $s$ symbols, and constructed $q$-ary
codes with $\log_2 n+O(1)$ bits of redundancy for fixed error parameters
and alphabet size. An interesting idea of burst-covering codes was suggested in~\cite{HiSc26}.
Finally, perfect codes in general were considered in~\cite{Etz22}.

The rest of the paper is organized as follows.
In Section~\ref{sec:ternary} a recursive construction is described for ternary perfect $2$-burst-correcting codes.
Conclusion and a short discussion is given in Section~\ref{sec:conclude}.

\section{Ternary Perfect $2$-Burst-Correcting Codes}
\label{sec:ternary}

A ternary perfect $2$-burst-correcting code of redundancy $r$ has length $n=\frac{3^r +3}{6}$.
The recursive construction starts with a parity-check matrix for code of length $n=\frac{3^r +3}{6}$ and
produces an $(r+3) \times (27n-13)$ parity-check matrix.
If $H=[h_1,\ldots,h_n]$, where $n=\frac{3^r +3}{6}$ is the parity-check matrix for the code with redundancy $r$,
we choose 27 blocks of length $n-1$ from $h_1,\ldots,h_{n-1}$ or $h_2,\cdots,h_n$ or $h_{n-1},\ldots,h_1$ or $h_n,\cdots,h_2$
and to each such blocks we add three appropriate rows. Between some of these 27 block we add one or two columns of length $r+3$
to form the missing syndromes of length $r+3$. There are $14$ such added columns.
The seeds for the construction are three parity-check matrices,
$5 \times 41$, $6 \times 122$, and $7 \times 365$.
These seeds for $r=5,6$ and $r=7$ are displayed in
Figs.~\ref{fig:seeds-five-six} and~\ref{fig:seed-seven}, respectively (there is no code for redundancy $4$ and length $14$).
The construction adapts the binary recursion to the four possible nonzero coefficient pairs.
The recursive construction distinguishes between even $n$ and odd $n$ of the $r \times n$ parity-check matrix.
The recursive matrices for even and odd input length $n$ appear in
Figs.~\ref{fig:recursion-even} and~\ref{fig:recursion-odd}, respectively.
These matrices will be called $H_\text{even}$ and $H_\text{odd}$, respectively.

To prove that the resulting code is a perfect $2$-burst-correcting code we have to show that for each syndrome $s$
obtained from $H$, there exists a syndrome $(s,t)$, where $t$ is a vector of length $3$ that can have each one of the
$27$ possibilities in $H_\text{even}$ and in $H_\text{odd}$. Moreover, also the syndrome $(0,t)$, where $0$ is a vector of length $r$
and $t$ is any nonzero ternary vector of length $3$, should be obtained.

In these two matrices, each block of length $n-1$ contains the $n-2$ consecutive columns $h_2,h_3,\ldots,h_{n-1}$.
The three rows below such $n-2$ columns form a $3 \times (n-2)$ matrix $\cA_k$, $1 \leq k \leq 27$,
whose columns are $a_2^k,a_3^k,\ldots,a_{n-1}^k$.
The matrix $\cA_k$ has period two in its columns, i.e., all the even columns ($h_i$, where $i$ is even) are the same and
all the odd columns are the same. All the columns below $h_2$ in $\cA_1,\cA_2,\ldots,\cA_{27}$
($a_2^1,a_2^2,\ldots,a_2^{27}$),
have each one of the $27$ ternary vectors of length three
exactly once. The same is true for all the columns below $h_3$ in $\cA_1,\cA_2,\ldots,\cA_{27}$.
Furthermore, each one of $a_2^1+a_3^1,a_2^2+a_3^2, \ldots, a_2^{27}+a_3^{27}$, also form each one
of the $27$ ternary vectors of length $3$ exactly once. The same is true for
$2a_2^1+2a_3^1,2a_2^2+2a_3^2, \ldots,2 a_2^{27}+2a_3^{27}$,
$2a_2^1+a_3^1,2a_2^2+a_3^2, \ldots, 2a_2^{27}+a_3^{27}$,
and $a_2^1+2a_3^1,a_2^2+2a_3^2, \ldots, a_2^{27}+2a_3^{27}$, where all the computations are modulo $3$.
This can also be observed by noting that if the three entries of the vector below $h_2$ is $(x,y,z)$ then
the vector below $h_3$ is $(z,x+z,y)$. A more formal proof will be given in the full-version of this paper.
Since the parity-check matrix $H$ form a perfect $2$-burst-correcting codes, it follows that
to prove that $H_\text{even}$ and $H_\text{odd}$ are parity-check matrices of perfect $2$-burst-correcting codes
we have to show that there are $27$ distinct ternary vectors of length $3$ below the combinations of columns,
$h_1$, $h_n$, $h_1+h_2$, $2h_1+h_2$, $h_{n-1}+h_n$, $2h_{n-1}+h_n$, and $26$ nonzero distinct such vectors for the zero column of length $r$.
This should be readily verified and hence using the seeds for $5 \leq r \leq 7$, we have the following consequence.

\begin{theorem}
For each $r \geq 5$ there exists an acyclic ternary linear perfect $2$-burst-correcting code of length $\frac{3^r +3}{6}$
and redundancy $r$.
\end{theorem}

\section{Conclusion}
\label{sec:conclude}

A recursive construction for ternary perfect $2$-burst-correcting codes was presented for
each redundancy $r \geq 5$. For redundancy $r=4$ such code does not exist.

Comprehensive search, but not exhaustive, on more than 10 sets of parameters, was not able to find any seed for other perfect $b$-burst-correcting
code. Hence, we suggest the following conjecture

\begin{conjecture}
Let $q$ be a prime power and let $1 < b <n$. If an acyclic linear code over $\F_q$ is perfect $b$-burst-correcting,
then $b=2$, $q \in \{2,3\}$, and its redundancy is at least 5.
\end{conjecture}

\begin{figure*}[p]
\centering
\scalebox{0.9}{\vbox{%
\begingroup
\fontsize{12}{14.5}\selectfont
\setlength{\arraycolsep}{0pt}
\renewcommand{\arraystretch}{1.0}
\seedtitle{Seed for \(r=7\): \(7\times 365\) matrix}

\seedstrip{Columns 1--73}{73}{[}{.}{%
  2 & 0 & 2 & 2 & 1 & 0 & 0 & 2 & 2 & 2 & 1 & 1 & 1 & 2 & 2 & 2 & 2 & 1 & 1 & 2 & 1 & 1 & 1 & 1 & 1 & 2 & 1 & 1 & 1 & 1 & 1 & 2 & 2 & 2 & 1 & 1 & 2 & 1 & 2 & 1 & 2 & 1 & 1 & 1 & 2 & 1 & 2 & 2 & 2 & 2 & 1 & 1 & 2 & 2 & 2 & 1 & 2 & 2 & 1 & 1 & 2 & 1 & 1 & 2 & 1 & 2 & 2 & 2 & 1 & 2 & 2 & 1 & 1\\
  1 & 1 & 2 & 1 & 2 & 1 & 0 & 1 & 2 & 0 & 0 & 0 & 0 & 1 & 0 & 0 & 0 & 0 & 0 & 0 & 1 & 2 & 2 & 1 & 0 & 0 & 1 & 0 & 1 & 0 & 1 & 0 & 1 & 2 & 1 & 2 & 0 & 0 & 2 & 1 & 0 & 1 & 1 & 0 & 0 & 0 & 1 & 2 & 1 & 0 & 1 & 1 & 1 & 2 & 2 & 0 & 2 & 2 & 2 & 1 & 1 & 0 & 1 & 0 & 2 & 2 & 0 & 2 & 0 & 1 & 0 & 0 & 1\\
  0 & 0 & 2 & 1 & 1 & 1 & 2 & 2 & 0 & 2 & 1 & 2 & 0 & 0 & 1 & 0 & 1 & 0 & 2 & 1 & 2 & 0 & 1 & 0 & 1 & 0 & 0 & 1 & 2 & 0 & 2 & 1 & 2 & 1 & 2 & 1 & 2 & 2 & 1 & 1 & 1 & 0 & 1 & 1 & 2 & 1 & 2 & 1 & 2 & 1 & 0 & 1 & 1 & 2 & 0 & 0 & 2 & 1 & 2 & 0 & 2 & 1 & 2 & 0 & 1 & 1 & 1 & 2 & 0 & 2 & 0 & 2 & 0\\
  0 & 0 & 0 & 0 & 0 & 0 & 0 & 0 & 0 & 1 & 2 & 1 & 0 & 1 & 0 & 2 & 0 & 1 & 2 & 0 & 2 & 2 & 0 & 2 & 0 & 0 & 1 & 0 & 1 & 1 & 1 & 1 & 2 & 1 & 2 & 2 & 2 & 0 & 2 & 1 & 1 & 1 & 1 & 2 & 0 & 1 & 1 & 1 & 2 & 0 & 0 & 0 & 1 & 0 & 1 & 2 & 1 & 2 & 0 & 2 & 2 & 1 & 2 & 1 & 2 & 2 & 2 & 1 & 2 & 1 & 1 & 1 & 2\\
  1 & 1 & 0 & 1 & 2 & 2 & 0 & 0 & 1 & 1 & 2 & 1 & 2 & 2 & 0 & 2 & 0 & 2 & 2 & 2 & 1 & 2 & 2 & 0 & 0 & 0 & 1 & 0 & 1 & 2 & 0 & 1 & 1 & 1 & 1 & 1 & 1 & 2 & 0 & 2 & 2 & 1 & 2 & 1 & 2 & 2 & 0 & 2 & 2 & 0 & 1 & 2 & 2 & 1 & 0 & 0 & 2 & 2 & 2 & 1 & 1 & 0 & 1 & 0 & 1 & 1 & 0 & 1 & 2 & 1 & 0 & 1 & 2\\
  1 & 1 & 1 & 1 & 1 & 1 & 1 & 1 & 0 & 2 & 2 & 0 & 1 & 0 & 0 & 2 & 1 & 2 & 1 & 1 & 0 & 2 & 2 & 0 & 0 & 1 & 1 & 2 & 0 & 1 & 1 & 1 & 0 & 1 & 0 & 1 & 0 & 0 & 1 & 2 & 0 & 0 & 0 & 0 & 2 & 2 & 0 & 0 & 1 & 1 & 2 & 1 & 0 & 0 & 1 & 0 & 0 & 1 & 1 & 0 & 1 & 0 & 1 & 2 & 0 & 1 & 0 & 1 & 2 & 1 & 1 & 1 & 1\\
  0 & 0 & 0 & 1 & 0 & 0 & 0 & 0 & 1 & 1 & 1 & 0 & 0 & 1 & 1 & 1 & 0 & 1 & 0 & 1 & 1 & 1 & 1 & 1 & 1 & 0 & 0 & 1 & 0 & 1 & 1 & 0 & 0 & 1 & 0 & 1 & 0 & 1 & 0 & 1 & 1 & 0 & 1 & 0 & 1 & 1 & 0 & 1 & 0 & 1 & 1 & 1 & 1 & 1 & 0 & 1 & 1 & 0 & 0 & 1 & 1 & 0 & 1 & 1 & 0 & 0 & 1 & 1 & 1 & 1 & 0 & 0 & 1%
}

\seedstrip{Columns 74--146}{73}{.}{.}{%
  1 & 2 & 2 & 2 & 1 & 1 & 2 & 1 & 2 & 1 & 2 & 2 & 1 & 1 & 1 & 2 & 2 & 1 & 1 & 2 & 2 & 1 & 1 & 2 & 1 & 2 & 2 & 1 & 2 & 1 & 2 & 1 & 1 & 1 & 1 & 1 & 2 & 1 & 2 & 1 & 2 & 2 & 1 & 1 & 1 & 2 & 2 & 2 & 1 & 2 & 2 & 2 & 1 & 1 & 2 & 1 & 2 & 1 & 1 & 2 & 2 & 2 & 1 & 2 & 1 & 2 & 2 & 2 & 2 & 1 & 2 & 2 & 2\\
  1 & 2 & 0 & 2 & 2 & 1 & 1 & 0 & 2 & 1 & 2 & 1 & 2 & 2 & 0 & 1 & 2 & 1 & 1 & 0 & 2 & 1 & 2 & 0 & 2 & 2 & 0 & 1 & 2 & 0 & 2 & 1 & 1 & 1 & 1 & 2 & 1 & 0 & 2 & 1 & 0 & 1 & 1 & 0 & 2 & 0 & 0 & 1 & 2 & 1 & 2 & 0 & 2 & 2 & 0 & 2 & 1 & 0 & 2 & 2 & 0 & 0 & 1 & 2 & 1 & 1 & 1 & 2 & 0 & 0 & 1 & 2 & 0\\
  1 & 1 & 0 & 1 & 2 & 0 & 2 & 0 & 2 & 0 & 2 & 0 & 2 & 2 & 1 & 0 & 2 & 0 & 2 & 1 & 1 & 2 & 1 & 1 & 1 & 0 & 0 & 0 & 1 & 2 & 2 & 2 & 1 & 1 & 2 & 0 & 1 & 0 & 1 & 0 & 2 & 0 & 0 & 2 & 1 & 2 & 0 & 2 & 0 & 0 & 0 & 2 & 2 & 1 & 0 & 1 & 2 & 2 & 0 & 1 & 1 & 2 & 1 & 1 & 1 & 0 & 0 & 2 & 1 & 1 & 2 & 0 & 0\\
  1 & 1 & 0 & 0 & 1 & 1 & 1 & 1 & 2 & 1 & 1 & 0 & 2 & 1 & 1 & 1 & 0 & 1 & 1 & 0 & 0 & 1 & 1 & 2 & 2 & 0 & 2 & 2 & 0 & 0 & 1 & 0 & 0 & 1 & 1 & 0 & 2 & 1 & 2 & 0 & 2 & 1 & 2 & 1 & 0 & 1 & 1 & 0 & 1 & 0 & 2 & 1 & 2 & 0 & 2 & 1 & 1 & 1 & 1 & 2 & 2 & 1 & 0 & 0 & 2 & 2 & 1 & 1 & 2 & 0 & 1 & 2 & 2\\
  0 & 1 & 2 & 2 & 1 & 1 & 2 & 0 & 0 & 1 & 0 & 0 & 0 & 1 & 2 & 2 & 2 & 2 & 1 & 1 & 0 & 2 & 2 & 1 & 1 & 1 & 0 & 0 & 2 & 1 & 0 & 0 & 0 & 1 & 0 & 1 & 1 & 2 & 0 & 2 & 0 & 0 & 2 & 2 & 0 & 2 & 1 & 1 & 1 & 1 & 0 & 1 & 1 & 1 & 1 & 0 & 0 & 0 & 2 & 1 & 2 & 0 & 0 & 1 & 1 & 0 & 0 & 0 & 2 & 0 & 1 & 1 & 2\\
  1 & 0 & 0 & 1 & 0 & 2 & 2 & 1 & 1 & 0 & 0 & 2 & 0 & 0 & 1 & 2 & 0 & 2 & 1 & 0 & 0 & 1 & 1 & 0 & 2 & 1 & 1 & 1 & 1 & 1 & 1 & 1 & 1 & 0 & 2 & 1 & 2 & 0 & 1 & 1 & 0 & 1 & 0 & 0 & 2 & 0 & 0 & 0 & 2 & 0 & 1 & 1 & 1 & 1 & 0 & 0 & 1 & 1 & 2 & 1 & 0 & 2 & 1 & 1 & 0 & 1 & 1 & 1 & 1 & 0 & 2 & 0 & 1\\
  1 & 0 & 1 & 1 & 0 & 1 & 1 & 0 & 0 & 1 & 0 & 1 & 1 & 1 & 0 & 1 & 1 & 1 & 0 & 1 & 0 & 0 & 0 & 0 & 1 & 0 & 1 & 0 & 0 & 0 & 0 & 0 & 0 & 0 & 1 & 0 & 1 & 1 & 1 & 1 & 1 & 1 & 1 & 1 & 1 & 1 & 0 & 1 & 1 & 1 & 0 & 0 & 0 & 1 & 1 & 0 & 1 & 0 & 1 & 1 & 1 & 1 & 1 & 1 & 1 & 1 & 0 & 1 & 0 & 0 & 1 & 1 & 0%
}

\seedstrip{Columns 147--219}{73}{.}{.}{%
  1 & 2 & 1 & 1 & 1 & 2 & 1 & 2 & 1 & 1 & 1 & 2 & 2 & 1 & 1 & 2 & 2 & 1 & 2 & 1 & 1 & 2 & 2 & 2 & 2 & 2 & 2 & 1 & 1 & 2 & 2 & 2 & 2 & 2 & 1 & 1 & 2 & 2 & 1 & 2 & 1 & 1 & 2 & 1 & 2 & 1 & 1 & 1 & 1 & 1 & 2 & 2 & 2 & 1 & 2 & 1 & 1 & 1 & 2 & 1 & 2 & 1 & 2 & 1 & 1 & 2 & 2 & 1 & 1 & 2 & 1 & 2 & 1\\
  2 & 2 & 1 & 1 & 0 & 2 & 0 & 0 & 0 & 0 & 1 & 2 & 2 & 0 & 0 & 0 & 1 & 1 & 0 & 1 & 0 & 2 & 2 & 1 & 2 & 0 & 0 & 0 & 0 & 1 & 0 & 2 & 0 & 2 & 0 & 1 & 0 & 0 & 1 & 2 & 1 & 0 & 1 & 1 & 0 & 0 & 0 & 1 & 2 & 1 & 0 & 1 & 0 & 0 & 2 & 2 & 2 & 0 & 0 & 1 & 1 & 0 & 0 & 0 & 2 & 0 & 2 & 1 & 2 & 1 & 0 & 2 & 2\\
  0 & 1 & 1 & 1 & 1 & 1 & 0 & 2 & 1 & 2 & 1 & 2 & 0 & 1 & 0 & 1 & 1 & 1 & 1 & 1 & 0 & 0 & 2 & 1 & 0 & 1 & 1 & 2 & 1 & 2 & 2 & 0 & 0 & 0 & 0 & 1 & 1 & 1 & 2 & 2 & 1 & 2 & 2 & 2 & 2 & 1 & 1 & 1 & 0 & 2 & 1 & 1 & 2 & 0 & 2 & 0 & 2 & 0 & 0 & 1 & 2 & 2 & 2 & 1 & 1 & 2 & 2 & 0 & 0 & 0 & 0 & 0 & 1\\
  2 & 1 & 2 & 2 & 2 & 2 & 2 & 0 & 2 & 0 & 2 & 0 & 0 & 0 & 1 & 1 & 2 & 0 & 2 & 1 & 0 & 1 & 1 & 2 & 0 & 1 & 2 & 1 & 2 & 1 & 1 & 0 & 0 & 0 & 2 & 1 & 1 & 1 & 2 & 1 & 1 & 0 & 1 & 0 & 2 & 1 & 1 & 1 & 2 & 2 & 2 & 0 & 1 & 0 & 0 & 2 & 0 & 0 & 2 & 0 & 1 & 2 & 0 & 0 & 1 & 0 & 1 & 1 & 2 & 2 & 1 & 0 & 1\\
  1 & 0 & 2 & 0 & 1 & 2 & 0 & 2 & 0 & 0 & 2 & 2 & 0 & 0 & 0 & 1 & 1 & 1 & 0 & 0 & 0 & 0 & 1 & 2 & 2 & 0 & 1 & 2 & 1 & 0 & 0 & 2 & 1 & 0 & 1 & 0 & 0 & 1 & 2 & 1 & 1 & 2 & 1 & 0 & 0 & 1 & 0 & 0 & 2 & 0 & 2 & 1 & 2 & 1 & 0 & 0 & 0 & 2 & 2 & 2 & 1 & 1 & 0 & 2 & 1 & 2 & 2 & 0 & 1 & 2 & 1 & 0 & 2\\
  0 & 1 & 0 & 1 & 1 & 1 & 1 & 1 & 2 & 0 & 2 & 1 & 1 & 1 & 2 & 2 & 0 & 1 & 2 & 0 & 2 & 2 & 2 & 1 & 1 & 1 & 2 & 1 & 2 & 0 & 0 & 0 & 1 & 1 & 0 & 2 & 1 & 1 & 0 & 0 & 1 & 2 & 1 & 2 & 2 & 1 & 1 & 1 & 1 & 2 & 2 & 2 & 2 & 0 & 2 & 2 & 2 & 2 & 0 & 0 & 0 & 0 & 0 & 2 & 2 & 1 & 1 & 1 & 1 & 1 & 0 & 0 & 2\\
  1 & 1 & 1 & 1 & 1 & 1 & 1 & 0 & 1 & 1 & 1 & 0 & 0 & 0 & 1 & 1 & 0 & 1 & 1 & 0 & 1 & 1 & 1 & 0 & 0 & 0 & 1 & 0 & 1 & 1 & 1 & 1 & 0 & 1 & 1 & 1 & 1 & 1 & 1 & 1 & 0 & 1 & 0 & 1 & 1 & 0 & 1 & 0 & 1 & 1 & 1 & 1 & 1 & 0 & 1 & 1 & 1 & 1 & 1 & 1 & 0 & 1 & 1 & 1 & 1 & 1 & 1 & 1 & 0 & 1 & 0 & 1 & 1%
}

\seedstrip{Columns 220--292}{73}{.}{.}{%
  2 & 2 & 1 & 1 & 2 & 1 & 2 & 2 & 2 & 2 & 1 & 1 & 2 & 2 & 1 & 1 & 2 & 1 & 2 & 1 & 2 & 1 & 2 & 1 & 2 & 1 & 2 & 1 & 2 & 2 & 1 & 2 & 2 & 1 & 1 & 1 & 2 & 1 & 2 & 2 & 2 & 1 & 1 & 1 & 2 & 1 & 1 & 1 & 2 & 1 & 1 & 1 & 2 & 1 & 2 & 2 & 2 & 2 & 2 & 2 & 1 & 2 & 2 & 2 & 1 & 1 & 1 & 2 & 2 & 1 & 2 & 2 & 2\\
  1 & 2 & 0 & 1 & 2 & 2 & 2 & 2 & 1 & 1 & 2 & 1 & 0 & 2 & 0 & 2 & 1 & 0 & 2 & 2 & 0 & 0 & 0 & 1 & 1 & 2 & 1 & 1 & 0 & 2 & 2 & 0 & 0 & 0 & 2 & 2 & 1 & 0 & 2 & 0 & 1 & 1 & 2 & 1 & 0 & 0 & 2 & 1 & 0 & 2 & 2 & 1 & 2 & 1 & 2 & 0 & 1 & 1 & 0 & 0 & 2 & 1 & 1 & 0 & 1 & 0 & 2 & 0 & 1 & 2 & 0 & 1 & 1\\
  0 & 1 & 0 & 0 & 1 & 1 & 0 & 1 & 0 & 0 & 2 & 0 & 0 & 1 & 2 & 2 & 2 & 1 & 0 & 1 & 1 & 1 & 1 & 1 & 0 & 1 & 2 & 1 & 1 & 1 & 2 & 0 & 2 & 1 & 1 & 0 & 1 & 0 & 2 & 0 & 2 & 0 & 0 & 0 & 0 & 2 & 2 & 1 & 1 & 1 & 0 & 2 & 1 & 2 & 0 & 0 & 0 & 1 & 0 & 0 & 2 & 2 & 0 & 0 & 2 & 1 & 0 & 1 & 2 & 2 & 0 & 0 & 0\\
  2 & 1 & 1 & 0 & 0 & 2 & 1 & 1 & 2 & 1 & 1 & 1 & 2 & 1 & 2 & 1 & 0 & 1 & 2 & 2 & 1 & 0 & 0 & 0 & 2 & 1 & 1 & 0 & 1 & 2 & 0 & 1 & 2 & 2 & 0 & 1 & 0 & 0 & 2 & 0 & 0 & 2 & 2 & 0 & 1 & 0 & 2 & 2 & 2 & 1 & 0 & 0 & 1 & 1 & 1 & 1 & 2 & 2 & 1 & 2 & 0 & 1 & 1 & 1 & 1 & 2 & 1 & 2 & 2 & 2 & 2 & 0 & 2\\
  1 & 1 & 1 & 2 & 0 & 0 & 2 & 1 & 0 & 2 & 0 & 2 & 1 & 0 & 0 & 2 & 1 & 1 & 0 & 1 & 0 & 2 & 0 & 2 & 0 & 0 & 1 & 1 & 2 & 1 & 1 & 2 & 0 & 0 & 1 & 1 & 2 & 0 & 1 & 0 & 2 & 1 & 0 & 0 & 0 & 2 & 2 & 2 & 0 & 2 & 2 & 1 & 0 & 0 & 1 & 0 & 2 & 2 & 1 & 1 & 0 & 2 & 2 & 2 & 2 & 1 & 2 & 1 & 1 & 2 & 1 & 2 & 0\\
  0 & 2 & 1 & 2 & 1 & 1 & 1 & 0 & 0 & 1 & 2 & 1 & 2 & 0 & 1 & 0 & 2 & 1 & 0 & 0 & 2 & 1 & 0 & 1 & 1 & 1 & 0 & 0 & 1 & 0 & 2 & 1 & 1 & 0 & 2 & 0 & 1 & 1 & 1 & 1 & 0 & 0 & 1 & 2 & 0 & 1 & 1 & 1 & 1 & 0 & 1 & 0 & 0 & 0 & 1 & 0 & 0 & 1 & 2 & 1 & 0 & 1 & 1 & 2 & 2 & 1 & 0 & 1 & 0 & 0 & 1 & 1 & 2\\
  1 & 1 & 1 & 1 & 0 & 0 & 1 & 1 & 1 & 0 & 1 & 0 & 1 & 1 & 1 & 1 & 1 & 1 & 1 & 1 & 1 & 1 & 0 & 0 & 0 & 1 & 1 & 0 & 0 & 1 & 1 & 0 & 0 & 1 & 1 & 1 & 1 & 1 & 1 & 1 & 1 & 0 & 1 & 1 & 1 & 1 & 1 & 1 & 0 & 1 & 0 & 0 & 0 & 1 & 1 & 0 & 1 & 1 & 1 & 1 & 1 & 0 & 1 & 1 & 1 & 0 & 1 & 0 & 1 & 1 & 0 & 1 & 1%
}

\seedstrip{Columns 293--365}{73}{.}{]}{%
  1 & 1 & 1 & 1 & 1 & 1 & 2 & 2 & 2 & 2 & 1 & 1 & 2 & 1 & 2 & 1 & 1 & 2 & 2 & 2 & 1 & 1 & 2 & 1 & 1 & 2 & 1 & 2 & 1 & 1 & 2 & 2 & 1 & 1 & 1 & 2 & 1 & 1 & 1 & 1 & 1 & 1 & 2 & 2 & 2 & 2 & 2 & 2 & 2 & 1 & 1 & 1 & 1 & 1 & 1 & 1 & 2 & 2 & 2 & 1 & 2 & 2 & 2 & 1 & 1 & 2 & 2 & 2 & 1 & 1 & 1 & 1 & 2\\
  0 & 0 & 2 & 0 & 0 & 0 & 2 & 1 & 0 & 2 & 0 & 1 & 2 & 2 & 2 & 0 & 1 & 1 & 0 & 1 & 2 & 2 & 2 & 1 & 1 & 1 & 2 & 2 & 0 & 0 & 1 & 0 & 2 & 2 & 0 & 1 & 0 & 1 & 1 & 0 & 1 & 1 & 2 & 1 & 0 & 0 & 1 & 1 & 1 & 2 & 1 & 2 & 1 & 1 & 2 & 0 & 0 & 1 & 1 & 0 & 2 & 2 & 1 & 2 & 1 & 0 & 2 & 0 & 1 & 0 & 2 & 1 & 1\\
  1 & 0 & 0 & 2 & 0 & 2 & 0 & 1 & 0 & 2 & 2 & 2 & 0 & 2 & 0 & 2 & 0 & 2 & 1 & 2 & 1 & 0 & 1 & 2 & 1 & 1 & 2 & 2 & 1 & 2 & 1 & 0 & 2 & 1 & 0 & 2 & 0 & 2 & 0 & 1 & 1 & 1 & 0 & 1 & 1 & 1 & 1 & 1 & 1 & 1 & 1 & 1 & 0 & 2 & 2 & 1 & 2 & 2 & 0 & 1 & 0 & 0 & 1 & 2 & 2 & 0 & 1 & 0 & 0 & 0 & 0 & 0 & 0\\
  1 & 0 & 2 & 0 & 0 & 1 & 1 & 2 & 0 & 2 & 0 & 1 & 2 & 0 & 0 & 1 & 0 & 2 & 1 & 1 & 1 & 0 & 1 & 2 & 0 & 1 & 1 & 0 & 2 & 0 & 0 & 0 & 1 & 1 & 2 & 2 & 2 & 2 & 2 & 1 & 0 & 2 & 0 & 1 & 2 & 0 & 1 & 0 & 0 & 2 & 0 & 0 & 1 & 2 & 1 & 0 & 0 & 1 & 2 & 1 & 1 & 1 & 2 & 1 & 0 & 1 & 2 & 0 & 2 & 1 & 1 & 0 & 0\\
  0 & 1 & 1 & 1 & 2 & 2 & 2 & 2 & 2 & 2 & 0 & 1 & 1 & 1 & 1 & 0 & 1 & 0 & 2 & 0 & 2 & 1 & 2 & 2 & 0 & 1 & 0 & 1 & 0 & 0 & 2 & 2 & 0 & 1 & 1 & 2 & 1 & 2 & 1 & 0 & 2 & 1 & 1 & 0 & 1 & 1 & 0 & 0 & 0 & 2 & 0 & 2 & 0 & 1 & 1 & 2 & 1 & 0 & 2 & 2 & 2 & 1 & 0 & 2 & 2 & 1 & 2 & 1 & 2 & 1 & 0 & 0 & 1\\
  2 & 0 & 0 & 2 & 0 & 2 & 1 & 2 & 2 & 1 & 2 & 2 & 0 & 1 & 0 & 1 & 0 & 0 & 2 & 2 & 1 & 1 & 1 & 1 & 0 & 2 & 1 & 1 & 1 & 1 & 2 & 1 & 1 & 0 & 0 & 0 & 2 & 1 & 2 & 1 & 2 & 1 & 2 & 0 & 1 & 2 & 1 & 1 & 0 & 0 & 0 & 1 & 0 & 1 & 2 & 0 & 2 & 1 & 1 & 0 & 2 & 0 & 1 & 1 & 0 & 1 & 2 & 0 & 1 & 0 & 0 & 1 & 2\\
  1 & 1 & 0 & 1 & 1 & 1 & 0 & 1 & 1 & 0 & 1 & 1 & 0 & 1 & 0 & 1 & 1 & 1 & 1 & 1 & 1 & 1 & 1 & 0 & 0 & 1 & 1 & 0 & 0 & 1 & 1 & 0 & 0 & 1 & 0 & 1 & 1 & 1 & 1 & 0 & 1 & 1 & 1 & 1 & 1 & 1 & 1 & 1 & 1 & 1 & 1 & 1 & 1 & 0 & 1 & 1 & 1 & 0 & 0 & 1 & 1 & 1 & 0 & 1 & 1 & 0 & 1 & 0 & 0 & 1 & 1 & 1 & 1%
}
\endgroup
}}\par
\caption{Seed parity-check matrix for $r=7$. Read the five strips from top to bottom.}
\label{fig:seed-seven}
\end{figure*}

% Full-width matrix figures retain the original 12-point notation.
\begin{figure*}[p]
\begin{minipage}{\textwidth}
\fontsize{12}{14.5}\selectfont
\setlength{\arraycolsep}{0pt}
\renewcommand{\arraystretch}{1.08}
\parityheader{Even \(n\)}

\diagramrow{*{7}{c}|c|*{7}{c}|*{7}{c}}{%
  \old{h_1} & \old{h_2} & \old{h_3} & \old{\cdots} & \old{h_{n-3}} & \old{h_{n-2}} & \old{h_{n-1}} & \old{h_n} & \old{h_n} & \old{h_{n-1}} & \old{h_{n-2}} & \old{h_{n-3}} & \old{\cdots} & \old{h_3} & \old{h_2} & \old{h_1} & \old{h_2} & \old{h_3} & \old{\cdots} & \old{h_{n-3}} & \old{h_{n-2}} & \old{h_{n-1}}\\ \hline
  0 & 0 & 1 & \cdots & 1 & 0 & 1 & 1 & 0 & 2 & 0 & 2 & \cdots & 2 & 0 & 1 & 2 & 2 & \cdots & 2 & 2 & 2\\
  0 & 2 & 1 & \cdots & 1 & 2 & 1 & 0 & 1 & 2 & 2 & 2 & \cdots & 2 & 2 & 2 & 1 & 1 & \cdots & 1 & 1 & 1\\
  0 & 1 & 2 & \cdots & 2 & 1 & 2 & 2 & 0 & 2 & 2 & 2 & \cdots & 2 & 2 & 0 & 2 & 1 & \cdots & 1 & 2 & 1%
}

\diagramrow{*{7}{c}|c|c|*{7}{c}|*{7}{c}}{%
  \old{h_n} & \old{h_{n-1}} & \old{h_{n-2}} & \old{h_{n-3}} & \old{\cdots} & \old{h_3} & \old{h_2} & \old{h_1} & \old{0} & \old{h_1} & \old{h_2} & \old{h_3} & \old{\cdots} & \old{h_{n-3}} & \old{h_{n-2}} & \old{h_{n-1}} & \old{h_n} & \old{h_{n-1}} & \old{h_{n-2}} & \old{h_{n-3}} & \old{\cdots} & \old{h_3} & \old{h_2}\\ \hline
  2 & 0 & 2 & 0 & \cdots & 0 & 2 & 2 & 1 & 2 & 1 & 1 & \cdots & 1 & 1 & 1 & 1 & 2 & 2 & 2 & \cdots & 2 & 2\\
  2 & 2 & 1 & 2 & \cdots & 2 & 1 & 1 & 1 & 0 & 2 & 2 & \cdots & 2 & 2 & 2 & 1 & 1 & 2 & 1 & \cdots & 1 & 2\\
  2 & 1 & 0 & 1 & \cdots & 1 & 0 & 2 & 0 & 2 & 1 & 2 & \cdots & 2 & 1 & 2 & 1 & 2 & 2 & 2 & \cdots & 2 & 2%
}

\diagramrow{*{7}{c}|*{7}{c}|c|*{7}{c}|c|c}{%
  \old{h_1} & \old{h_2} & \old{h_3} & \old{\cdots} & \old{h_{n-3}} & \old{h_{n-2}} & \old{h_{n-1}} & \old{h_n} & \old{h_{n-1}} & \old{h_{n-2}} & \old{h_{n-3}} & \old{\cdots} & \old{h_3} & \old{h_2} & \old{h_1} & \old{h_2} & \old{h_3} & \old{\cdots} & \old{h_{n-3}} & \old{h_{n-2}} & \old{h_{n-1}} & \old{h_n} & \old{0} & \old{0}\\ \hline
  2 & 0 & 2 & \cdots & 2 & 0 & 2 & 0 & 2 & 0 & 2 & \cdots & 2 & 0 & 1 & 1 & 1 & \cdots & 1 & 1 & 1 & 2 & 0 & 0\\
  1 & 1 & 2 & \cdots & 2 & 1 & 2 & 0 & 2 & 0 & 2 & \cdots & 2 & 0 & 1 & 1 & 2 & \cdots & 2 & 1 & 2 & 0 & 1 & 1\\
  1 & 2 & 1 & \cdots & 1 & 2 & 1 & 2 & 0 & 2 & 0 & \cdots & 0 & 2 & 1 & 1 & 1 & \cdots & 1 & 1 & 1 & 1 & 1 & 0%
}

\diagramrow{*{7}{c}|c|*{7}{c}|*{7}{c}|c}{%
  \old{h_n} & \old{h_{n-1}} & \old{h_{n-2}} & \old{h_{n-3}} & \old{\cdots} & \old{h_3} & \old{h_2} & \old{h_1} & \old{h_1} & \old{h_2} & \old{h_3} & \old{\cdots} & \old{h_{n-3}} & \old{h_{n-2}} & \old{h_{n-1}} & \old{h_n} & \old{h_{n-1}} & \old{h_{n-2}} & \old{h_{n-3}} & \old{\cdots} & \old{h_3} & \old{h_2} & \old{h_1}\\ \hline
  1 & 0 & 0 & 0 & \cdots & 0 & 0 & 0 & 2 & 0 & 0 & \cdots & 0 & 0 & 0 & 0 & 1 & 0 & 1 & \cdots & 1 & 0 & 1\\
  2 & 0 & 0 & 0 & \cdots & 0 & 0 & 0 & 2 & 1 & 0 & \cdots & 0 & 1 & 0 & 1 & 1 & 1 & 1 & \cdots & 1 & 1 & 2\\
  0 & 0 & 0 & 0 & \cdots & 0 & 0 & 1 & 0 & 0 & 1 & \cdots & 1 & 0 & 1 & 1 & 1 & 1 & 1 & \cdots & 1 & 1 & 1%
}

\diagramrow{*{7}{c}|c|*{7}{c}|*{7}{c}}{%
  \old{h_1} & \old{h_2} & \old{h_3} & \old{\cdots} & \old{h_{n-3}} & \old{h_{n-2}} & \old{h_{n-1}} & \old{h_n} & \old{h_n} & \old{h_{n-1}} & \old{h_{n-2}} & \old{h_{n-3}} & \old{\cdots} & \old{h_3} & \old{h_2} & \old{h_1} & \old{h_2} & \old{h_3} & \old{\cdots} & \old{h_{n-3}} & \old{h_{n-2}} & \old{h_{n-1}}\\ \hline
  0 & 2 & 1 & \cdots & 1 & 2 & 1 & 1 & 2 & 0 & 0 & 0 & \cdots & 0 & 0 & 2 & 1 & 2 & \cdots & 2 & 1 & 2\\
  2 & 0 & 0 & \cdots & 0 & 0 & 0 & 0 & 1 & 0 & 2 & 0 & \cdots & 0 & 2 & 2 & 2 & 0 & \cdots & 0 & 2 & 0\\
  0 & 1 & 0 & \cdots & 0 & 1 & 0 & 1 & 0 & 2 & 0 & 2 & \cdots & 2 & 0 & 1 & 2 & 2 & \cdots & 2 & 2 & 2%
}

\diagramrow{*{7}{c}|c|*{7}{c}|*{7}{c}}{%
  \old{h_n} & \old{h_{n-1}} & \old{h_{n-2}} & \old{h_{n-3}} & \old{\cdots} & \old{h_3} & \old{h_2} & \old{h_1} & \old{h_1} & \old{h_2} & \old{h_3} & \old{\cdots} & \old{h_{n-3}} & \old{h_{n-2}} & \old{h_{n-1}} & \old{h_n} & \old{h_{n-1}} & \old{h_{n-2}} & \old{h_{n-3}} & \old{\cdots} & \old{h_3} & \old{h_2}\\ \hline
  2 & 1 & 2 & 1 & \cdots & 1 & 2 & 1 & 0 & 2 & 2 & \cdots & 2 & 2 & 2 & 1 & 0 & 2 & 0 & \cdots & 0 & 2\\
  0 & 0 & 1 & 0 & \cdots & 0 & 1 & 0 & 0 & 0 & 1 & \cdots & 1 & 0 & 1 & 2 & 2 & 0 & 2 & \cdots & 2 & 0\\
  2 & 1 & 1 & 1 & \cdots & 1 & 1 & 1 & 2 & 2 & 0 & \cdots & 0 & 2 & 0 & 2 & 0 & 0 & 0 & \cdots & 0 & 0%
}

\diagramrow{*{7}{c}|*{7}{c}|*{7}{c}|c}{%
  \old{h_1} & \old{h_2} & \old{h_3} & \old{\cdots} & \old{h_{n-3}} & \old{h_{n-2}} & \old{h_{n-1}} & \old{h_n} & \old{h_{n-1}} & \old{h_{n-2}} & \old{h_{n-3}} & \old{\cdots} & \old{h_3} & \old{h_2} & \old{h_1} & \old{h_2} & \old{h_3} & \old{\cdots} & \old{h_{n-3}} & \old{h_{n-2}} & \old{h_{n-1}} & \old{h_n}\\ \hline
  1 & 1 & 2 & \cdots & 2 & 1 & 2 & 0 & 0 & 1 & 0 & \cdots & 0 & 1 & 2 & 2 & 0 & \cdots & 0 & 2 & 0 & 0\\
  0 & 1 & 0 & \cdots & 0 & 1 & 0 & 2 & 1 & 1 & 1 & \cdots & 1 & 1 & 0 & 2 & 2 & \cdots & 2 & 2 & 2 & 2\\
  0 & 2 & 1 & \cdots & 1 & 2 & 1 & 0 & 1 & 0 & 1 & \cdots & 1 & 0 & 1 & 0 & 2 & \cdots & 2 & 0 & 2 & 1%
}

\diagramrow{*{7}{c}|*{7}{c}|*{7}{c}|c}{%
  \old{h_n} & \old{h_{n-1}} & \old{h_{n-2}} & \old{h_{n-3}} & \old{\cdots} & \old{h_3} & \old{h_2} & \old{h_1} & \old{h_2} & \old{h_3} & \old{\cdots} & \old{h_{n-3}} & \old{h_{n-2}} & \old{h_{n-1}} & \old{h_n} & \old{h_{n-1}} & \old{h_{n-2}} & \old{h_{n-3}} & \old{\cdots} & \old{h_3} & \old{h_2} & \old{h_1}\\ \hline
  1 & 0 & 1 & 0 & \cdots & 0 & 1 & 0 & 2 & 1 & \cdots & 1 & 2 & 1 & 0 & 1 & 1 & 1 & \cdots & 1 & 1 & 0\\
  1 & 1 & 2 & 1 & \cdots & 1 & 2 & 1 & 2 & 0 & \cdots & 0 & 2 & 0 & 1 & 2 & 0 & 2 & \cdots & 2 & 0 & 1\\
  2 & 2 & 0 & 2 & \cdots & 2 & 0 & 1 & 1 & 2 & \cdots & 2 & 1 & 2 & 2 & 0 & 1 & 0 & \cdots & 0 & 1 & 0%
}

\diagramrow{*{7}{c}|c|*{7}{c}|c|*{7}{c}}{%
  \old{h_1} & \old{h_2} & \old{h_3} & \old{\cdots} & \old{h_{n-3}} & \old{h_{n-2}} & \old{h_{n-1}} & \old{h_n} & \old{h_n} & \old{h_{n-1}} & \old{h_{n-2}} & \old{h_{n-3}} & \old{\cdots} & \old{h_3} & \old{h_2} & \old{h_1} & \old{h_2} & \old{h_3} & \old{\cdots} & \old{h_{n-3}} & \old{h_{n-2}} & \old{h_{n-1}} & \old{h_n}\\ \hline
  1 & 1 & 0 & \cdots & 0 & 1 & 0 & 0 & 1 & 2 & 1 & 2 & \cdots & 2 & 1 & 0 & 0 & 1 & \cdots & 1 & 0 & 1 & 0\\
  1 & 0 & 1 & \cdots & 1 & 0 & 1 & 0 & 2 & 0 & 0 & 0 & \cdots & 0 & 0 & 2 & 0 & 1 & \cdots & 1 & 0 & 1 & 0\\
  0 & 0 & 0 & \cdots & 0 & 0 & 0 & 1 & 1 & 0 & 2 & 0 & \cdots & 0 & 2 & 1 & 1 & 0 & \cdots & 0 & 1 & 0 & 0%
}

\end{minipage}
\caption{Recursive parity-check matrix for even input length $n$. Read the strips from top to bottom.}
\label{fig:recursion-even}
\end{figure*}
\par

\begin{figure*}[p]
\begin{minipage}{\textwidth}
\fontsize{12}{14.5}\selectfont
\setlength{\arraycolsep}{0pt}
\renewcommand{\arraystretch}{1.08}
\parityheader{Odd \(n\)}

\diagramrow{*{7}{c}|*{7}{c}|*{7}{c}|c}{%
  \old{h_1} & \old{h_2} & \old{h_3} & \old{\cdots} & \old{h_{n-3}} & \old{h_{n-2}} & \old{h_{n-1}} & \old{h_n} & \old{h_{n-1}} & \old{h_{n-2}} & \old{h_{n-3}} & \old{\cdots} & \old{h_3} & \old{h_2} & \old{h_1} & \old{h_2} & \old{h_3} & \old{\cdots} & \old{h_{n-3}} & \old{h_{n-2}} & \old{h_{n-1}} & \old{h_n}\\ \hline
  0 & 0 & 1 & \cdots & 0 & 1 & 0 & 2 & 1 & 2 & 1 & \cdots & 2 & 1 & 2 & 0 & 0 & \cdots & 0 & 0 & 0 & 1\\
  0 & 2 & 1 & \cdots & 2 & 1 & 2 & 0 & 2 & 0 & 2 & \cdots & 0 & 2 & 2 & 2 & 0 & \cdots & 2 & 0 & 2 & 0\\
  0 & 1 & 2 & \cdots & 1 & 2 & 1 & 2 & 2 & 2 & 2 & \cdots & 2 & 2 & 1 & 0 & 2 & \cdots & 0 & 2 & 0 & 1%
}

\diagramrow{*{7}{c}|*{7}{c}|*{7}{c}}{%
  \old{h_n} & \old{h_{n-1}} & \old{h_{n-2}} & \old{h_{n-3}} & \old{\cdots} & \old{h_3} & \old{h_2} & \old{h_1} & \old{h_2} & \old{h_3} & \old{\cdots} & \old{h_{n-3}} & \old{h_{n-2}} & \old{h_{n-1}} & \old{h_n} & \old{h_{n-1}} & \old{h_{n-2}} & \old{h_{n-3}} & \old{\cdots} & \old{h_3} & \old{h_2}\\ \hline
  2 & 2 & 2 & 2 & \cdots & 2 & 2 & 2 & 0 & 2 & \cdots & 0 & 2 & 0 & 1 & 1 & 2 & 1 & \cdots & 2 & 1\\
  1 & 2 & 1 & 2 & \cdots & 1 & 2 & 1 & 1 & 2 & \cdots & 1 & 2 & 1 & 1 & 0 & 0 & 0 & \cdots & 0 & 0\\
  0 & 2 & 2 & 2 & \cdots & 2 & 2 & 1 & 2 & 1 & \cdots & 2 & 1 & 2 & 1 & 2 & 0 & 2 & \cdots & 0 & 2%
}

\diagramrow{*{7}{c}|*{7}{c}|*{7}{c}|c}{%
  \old{h_1} & \old{h_2} & \old{h_3} & \old{\cdots} & \old{h_{n-3}} & \old{h_{n-2}} & \old{h_{n-1}} & \old{h_n} & \old{h_{n-1}} & \old{h_{n-2}} & \old{h_{n-3}} & \old{\cdots} & \old{h_3} & \old{h_2} & \old{h_1} & \old{h_2} & \old{h_3} & \old{\cdots} & \old{h_{n-3}} & \old{h_{n-2}} & \old{h_{n-1}} & \old{h_n}\\ \hline
  0 & 0 & 1 & \cdots & 0 & 1 & 0 & 0 & 2 & 1 & 2 & \cdots & 1 & 2 & 0 & 1 & 0 & \cdots & 1 & 0 & 1 & 1\\
  2 & 0 & 1 & \cdots & 0 & 1 & 0 & 1 & 2 & 0 & 2 & \cdots & 0 & 2 & 1 & 2 & 1 & \cdots & 2 & 1 & 2 & 2\\
  1 & 1 & 0 & \cdots & 1 & 0 & 1 & 1 & 1 & 2 & 1 & \cdots & 2 & 1 & 1 & 0 & 2 & \cdots & 0 & 2 & 0 & 1%
}

\diagramrow{*{7}{c}|*{7}{c}|c|*{7}{c}}{%
  \old{h_n} & \old{h_{n-1}} & \old{h_{n-2}} & \old{h_{n-3}} & \old{\cdots} & \old{h_3} & \old{h_2} & \old{h_1} & \old{h_2} & \old{h_3} & \old{\cdots} & \old{h_{n-3}} & \old{h_{n-2}} & \old{h_{n-1}} & \old{h_n} & \old{h_n} & \old{h_{n-1}} & \old{h_{n-2}} & \old{h_{n-3}} & \old{\cdots} & \old{h_3} & \old{h_2}\\ \hline
  0 & 2 & 0 & 2 & \cdots & 0 & 2 & 1 & 1 & 2 & \cdots & 1 & 2 & 1 & 0 & 1 & 0 & 2 & 0 & \cdots & 2 & 0\\
  0 & 0 & 2 & 0 & \cdots & 2 & 0 & 0 & 1 & 0 & \cdots & 1 & 0 & 1 & 1 & 0 & 0 & 2 & 0 & \cdots & 2 & 0\\
  1 & 0 & 0 & 0 & \cdots & 0 & 0 & 0 & 2 & 1 & \cdots & 2 & 1 & 2 & 0 & 2 & 2 & 0 & 2 & \cdots & 0 & 2%
}

\diagramrow{*{7}{c}|*{7}{c}|c|*{7}{c}|c|c}{%
  \old{h_1} & \old{h_2} & \old{h_3} & \old{\cdots} & \old{h_{n-3}} & \old{h_{n-2}} & \old{h_{n-1}} & \old{h_n} & \old{h_{n-1}} & \old{h_{n-2}} & \old{h_{n-3}} & \old{\cdots} & \old{h_3} & \old{h_2} & \old{h_1} & \old{h_2} & \old{h_3} & \old{\cdots} & \old{h_{n-3}} & \old{h_{n-2}} & \old{h_{n-1}} & \old{h_n} & \old{0} & \old{0}\\ \hline
  1 & 1 & 1 & \cdots & 1 & 1 & 1 & 0 & 1 & 0 & 1 & \cdots & 0 & 1 & 2 & 2 & 0 & \cdots & 2 & 0 & 2 & 1 & 0 & 0\\
  1 & 1 & 2 & \cdots & 1 & 2 & 1 & 0 & 1 & 1 & 1 & \cdots & 1 & 1 & 0 & 2 & 2 & \cdots & 2 & 2 & 2 & 2 & 1 & 1\\
  1 & 1 & 1 & \cdots & 1 & 1 & 1 & 2 & 0 & 1 & 0 & \cdots & 1 & 0 & 1 & 0 & 2 & \cdots & 0 & 2 & 0 & 0 & 0 & 1%
}

\diagramrow{*{7}{c}|c|*{7}{c}|*{7}{c}|c|c}{%
  \old{h_n} & \old{h_{n-1}} & \old{h_{n-2}} & \old{h_{n-3}} & \old{\cdots} & \old{h_3} & \old{h_2} & \old{h_1} & \old{h_1} & \old{h_2} & \old{h_3} & \old{\cdots} & \old{h_{n-3}} & \old{h_{n-2}} & \old{h_{n-1}} & \old{h_n} & \old{h_{n-1}} & \old{h_{n-2}} & \old{h_{n-3}} & \old{\cdots} & \old{h_3} & \old{h_2} & \old{h_1} & \old{0}\\ \hline
  2 & 0 & 1 & 0 & \cdots & 1 & 0 & 1 & 0 & 2 & 1 & \cdots & 2 & 1 & 2 & 0 & 1 & 1 & 1 & \cdots & 1 & 1 & 2 & 1\\
  0 & 1 & 1 & 1 & \cdots & 1 & 1 & 2 & 2 & 0 & 0 & \cdots & 0 & 0 & 0 & 2 & 2 & 2 & 2 & \cdots & 2 & 2 & 0 & 1\\
  1 & 1 & 1 & 1 & \cdots & 1 & 1 & 1 & 0 & 1 & 0 & \cdots & 1 & 0 & 1 & 0 & 1 & 2 & 1 & \cdots & 2 & 1 & 2 & 0%
}

\diagramrow{*{7}{c}|*{7}{c}|c|*{7}{c}}{%
  \old{h_1} & \old{h_2} & \old{h_3} & \old{\cdots} & \old{h_{n-3}} & \old{h_{n-2}} & \old{h_{n-1}} & \old{h_n} & \old{h_{n-1}} & \old{h_{n-2}} & \old{h_{n-3}} & \old{\cdots} & \old{h_3} & \old{h_2} & \old{h_1} & \old{h_1} & \old{h_2} & \old{h_3} & \old{\cdots} & \old{h_{n-3}} & \old{h_{n-2}} & \old{h_{n-1}}\\ \hline
  2 & 2 & 0 & \cdots & 2 & 0 & 2 & 1 & 1 & 0 & 1 & \cdots & 0 & 1 & 1 & 0 & 1 & 1 & \cdots & 1 & 1 & 1\\
  1 & 1 & 2 & \cdots & 1 & 2 & 1 & 2 & 0 & 1 & 0 & \cdots & 1 & 0 & 1 & 1 & 0 & 2 & \cdots & 0 & 2 & 0\\
  2 & 0 & 1 & \cdots & 0 & 1 & 0 & 2 & 0 & 0 & 0 & \cdots & 0 & 0 & 0 & 0 & 1 & 0 & \cdots & 1 & 0 & 1%
}

\diagramrow{*{7}{c}|c|*{7}{c}|c|*{7}{c}}{%
  \old{h_n} & \old{h_{n-1}} & \old{h_{n-2}} & \old{h_{n-3}} & \old{\cdots} & \old{h_3} & \old{h_2} & \old{h_1} & \old{h_1} & \old{h_2} & \old{h_3} & \old{\cdots} & \old{h_{n-3}} & \old{h_{n-2}} & \old{h_{n-1}} & \old{h_n} & \old{h_n} & \old{h_{n-1}} & \old{h_{n-2}} & \old{h_{n-3}} & \old{\cdots} & \old{h_3} & \old{h_2}\\ \hline
  2 & 2 & 1 & 2 & \cdots & 1 & 2 & 1 & 0 & 2 & 2 & \cdots & 2 & 2 & 2 & 1 & 0 & 2 & 2 & 2 & \cdots & 2 & 2\\
  2 & 1 & 0 & 1 & \cdots & 0 & 1 & 0 & 0 & 0 & 1 & \cdots & 0 & 1 & 0 & 1 & 2 & 1 & 1 & 1 & \cdots & 1 & 1\\
  2 & 1 & 1 & 1 & \cdots & 1 & 1 & 1 & 2 & 2 & 0 & \cdots & 2 & 0 & 2 & 2 & 1 & 2 & 1 & 2 & \cdots & 1 & 2%
}

\diagramrow{*{7}{c}|*{7}{c}|c|c|*{7}{c}}{%
  \old{h_1} & \old{h_2} & \old{h_3} & \old{\cdots} & \old{h_{n-3}} & \old{h_{n-2}} & \old{h_{n-1}} & \old{h_n} & \old{h_{n-1}} & \old{h_{n-2}} & \old{h_{n-3}} & \old{\cdots} & \old{h_3} & \old{h_2} & \old{h_1} & \old{h_1} & \old{h_2} & \old{h_3} & \old{\cdots} & \old{h_{n-3}} & \old{h_{n-2}} & \old{h_{n-1}} & \old{h_n}\\ \hline
  1 & 0 & 2 & \cdots & 0 & 2 & 0 & 0 & 0 & 0 & 0 & \cdots & 0 & 0 & 2 & 0 & 0 & 0 & \cdots & 0 & 0 & 0 & 0\\
  2 & 2 & 2 & \cdots & 2 & 2 & 2 & 1 & 1 & 0 & 1 & \cdots & 0 & 1 & 2 & 0 & 0 & 0 & \cdots & 0 & 0 & 0 & 0\\
  0 & 2 & 2 & \cdots & 2 & 2 & 2 & 2 & 0 & 1 & 0 & \cdots & 1 & 0 & 0 & 1 & 0 & 0 & \cdots & 0 & 0 & 0 & 0%
}

\end{minipage}
\caption{Recursive parity-check matrix for odd input length $n$. Read the strips from top to bottom.}
\label{fig:recursion-odd}
\end{figure*}

%%%%%%%%%%%%%%%%%%%%%%%%%%%%%%%%%%%%%%%%%%%%%%%%%%%%%%%%%%%%%%%%%%%%%%
%%%%%%%%%%%%%%%%%%%%%%%%%%%%%%%%%%%%%%%%%%%%%%%%%%%%%%%%%%%%%%%%%%%%%%
%%%%%%%%%%%%%%%%%%%%%%%%%%%%%%%%%%%%%%%%%%%%%%%%%%%%%%%%%%%%%%%%%%%%%%
%%%%%%%%%%%%%%%%%%%%%%%%%%%%%%%%%%%%%%%%%%%%%%%%%%%%%%%%%%%%%%%%%%%%%%

% Finish all matrix figures before the separate final references page.
\clearpage
\IEEEtriggeratref{7}

\end{document}